\documentstyle[psfig,epsf,amssymb]{aa}
\begin{document}

\thesaurus{02(02.01.2)}
\title{The global structure of thin, stratified $\alpha$-discs\\ and the reliability of the one layer approximation}
\titlerunning{On the global structure of thin, stratified $\alpha$-discs}
             \author{Jean-Marc Hur\'e$^{1,2}$ and Fr\'ed\'eric Galliano$^{1,}$\thanks{Present address: SAp/CEA/Saclay, L'Orme des Merisiers, 91191 Gif-sur-Yvette, France}}
             \offprints{Jean-Marc.Hure@obspm.fr}
             \institute{$^1$DAEC et UMR 8631 du CNRS, Observatoire de Paris-Meudon, Place Jules Janssen, 92195 Meudon Cedex, France\\
                        $^2$Universit\'e Paris 7 Denis Diderot, 2 Place Jussieu, 75251 Paris Cedex 05, France}
             \date{Received 12 July 2000; accepted 19 October 2000}
             \maketitle

\begin{abstract}

We report the results of a systematic comparison between the vertically
averaged model and the vertically explicit model of steady state, Keplerian, optically thick $\alpha$-discs. The simulations have concerned discs currently found in three
 different systems: dwarf novae, young stellar objects and active galactic nuclei. In each case, we have explored four decades of accretion rates and almost the whole disc area (except the narrow region where the vertically averaged model has degenerate solutions). We find that the one layer approach gives a remarkably good estimate of the main physical quantities in the disc, and specially the temperature at the equatorial plane which is accurate to within 30 $\%$ for
cases considered. The major deviations (by a factor $\lesssim 4$) are observed on the disc half-thickness.
 The sensitivity of the results to the $\alpha$-parameter value has been tested for $0.001 \le \alpha \le 0.1$ and appears to be weak. This study suggests that the ``precision'' of the vertically averaged model which is easy to implement should be sufficient in practice for many astrophysical applications.


\end{abstract}

\section{Introduction}

Recent investigations on MHD turbulence have confirmed the existence of a mechanism able to extract angular momentum from weakly magnetized accretion discs (Balbus \& Hawley, 1991; Fleming, Stone \& Hawley, 2000; Miller \& Stone, 2000). Many aspects of accretion are, however, still not understood. In the meanwhile, present disc models mostly incorporate the $\alpha$-prescription (Shakura \& Sunyaev, 1973) and are of increasing complexity. For instance, models of steady state, Keplerian, vertically stratified $\alpha$-discs (e.g. Cannizzo, 1992; Hameury et al., 1998; D'Alessio et al., 1999) should definitely supplant the more widespread, vertically averaged version (Pringle, 1981) which, despite a great simplicity and flexibility, cannot account for fine physical effects, remains in essence somewhat limited and contains some artefacts, like degenerate solutions (Collin-Souffrin \& Dumont, 1990).
Vertically explicit disc models offer in principle a much better reliability but, due to the uncertainty on the $\alpha$-formalism, they require an extra prescription specifying the depth-dependence of the viscous energy release (Meyer \& Meyer-Hofmeister, 1982; Mineshige \& Osaki, 1983; Cannizzo \& Cameron, 1988; Hameury et al., 1998). From a numerical point of view, the bi-dimensional problem which involves partial/ordinary differential equations is much less trivial to solve than the vertically averaged problem, also meaning a much larger computational time (by a factor $\sim 100-1000$ say, corresponding to the typical resolution in the $z$-direction). It is therefore legitimate to ask, from a {\it quantitative point of view} at least, which benefits the vertically explicit model really brings. The aim of this report is to show how, for steady Keplerian $\alpha$-discs, the one layer approach and the vertically explicit approach compare. The main motivation is both to seek where the differences appear in an $\alpha$-disc and to quantify these as functions of the accretion rate and value of the viscosity parameter. This investigation concerns three generic systems containing a disc: an active galactic nucleus, a young stellar object and a dwarf nova.  
\begin{figure*}[htbp]
    \psfig{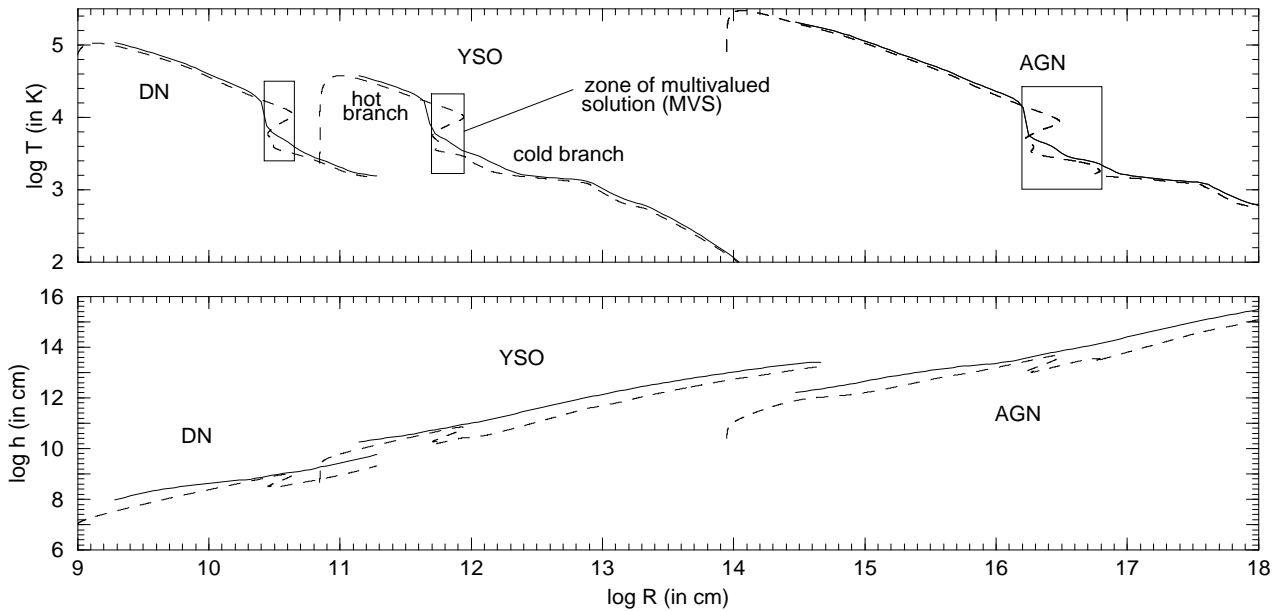}
  \caption{Midplane temperature
    ({\it top}) and disc half-thickness ({\it bottom}) versus the radius for
     $M=0.61$ M$_\odot$, $\dot{M}=10^{-10}$ M$_\odot$/yr and
     $\alpha=10^{-2}$ (DN; {\it left}), $M=1$ M$_\odot$, $\dot{M}=10^{-7}$ M$_\odot$/yr and $\alpha=10^{-2}$ (YSO; {\it middle}) and
     $M=10^8$ M$_\odot$, $\dot{M}=10^{-2}$ M$_\odot$/yr and
     $\alpha=0.1$ (AGN; {\it right}). The systematic
     comparisons between the VA-model ({\it dashed lines}) and the VE-model ({\it solid lines}) concern the hot and cold branches, outside the zone where the VA-model has a multivalued solution (MVS). }
  \label{courbe}
\end{figure*}

\section{Vertically explicit vs. vertically averaged models: hypothesis and restrictions} \label{models}

We compare the structure of steady state Keplerian accretion discs computed from the vertically averaged model (hereafter, the VA-model) and from the vertically explicit model (hereafter, the VE-model). The difference between the two resides therefore in the resolution of the vertical stratification. This analysis is carried out in the framework of the $\alpha$-theory (Shakura \& Sunyaev, 1973), with a uniform $\alpha$-parameter value. For a detailed description of the models, the reader is referred to Hur\'e (1998, 2000).

In order to give a meaning to the comparison, the two models are made as close as possible and reduced to their simplest form. In particular, we use the same Rosseland and Planck opacity grids and equation of state corresponding to a solar mix at LTE for both models. Convective transport is not taken into account in the VE-model, for several reasons. First, it cannot be included in the one layer version. Second, it plays a role in regions of high opacity, namely where hydrogen recombines. In the framework of the Mixing Length Theory (e.g. Cox \& Giuli, 1968), the disc region mainly involved has a small size and convection does not drastically change the disc structure (Meyer \& Meyer-Hofmeister, 1982; Cannizzo, 1993). Note that the weak effect of convection is sustained in the work by Gu, Vishniac \& Cannizzo (1999). Third, as explained below, we shall not analyze the region where the one layer has multiple solutions and which precisely encompasses the convectively unstable zone.

Also left aside is external irradiation. It depends on many parameters and has little effect on the global quantities (midplane temperature, surface density, etc.) as long as the disc remains optically thick.
Regarding the depth-dependent viscosity prescription, we adopt the most common parameterization of the heat flux gradient, namely
\begin{equation}
\frac{dF}{dz} = \frac{3}{2} \alpha P \Omega_{\rm K}
\end{equation}

where  $P$ is here the {\it total pressure} and $\Omega_{\rm K}$ is the Keplerian angular velocity. We also neglect the disc self-gravity.

\section{Results} \label{results}

The disc structure depends mainly on three input parameters: the central mass $M$, the accretion rate $\dot{M}$ and value of the $\alpha$-viscosity parameter. The departures between the VA-model and the VE-model are measured {\it at the same radius} $R$ by the ratio
\begin{equation} \label{eqn:epsi}
\epsilon(x) = \frac{x^{\rm VE-model}}{x^{\rm VA-model}}
\label{eq:eps}
\end{equation}

where $x$ is one of the following quantities: the
temperature $T_{\rm mid}$ (the subscript ``mid'' refers to the midplane), half the disc thickness $h$ (taken as the altitude of the photosphere's base in the VE-model and as the pressure scale height in the VA-model), the density $\rho_{\rm mid}$,
the total surface density $\Sigma_{\rm t} = \int_{-\infty}^{+\infty}{\rho dz}$
and the optical thickness $\tau_{\rm t} = 2 \int_{0}^{+\infty}{\kappa
  \rho dz}$. Note that $\epsilon$ is a function of the radius.

\begin{figure}[htbp]
    \psfig{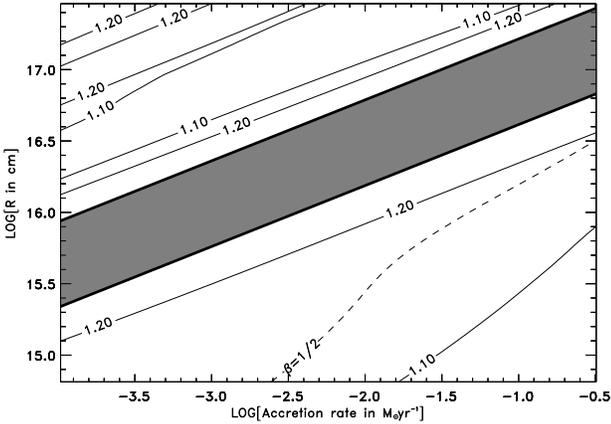}
  \caption{Iso-values of $\epsilon (T_{\rm mid})$ in the $(\log R, \log \dot{M})-$plane, in the AGN case
     ($M=10^8$ M$_\odot$ and $\alpha=0.01$). The MVS region (see Fig. \ref{courbe} and Eq.(\ref{eq:mvsregion})) is shown in grey. The dashed line labeled $\beta=\frac{1}{2}$ marks the limit between the gas pressure supported disc (at large radii) and the radiation pressure supported disc.}
  \label{fig:htrmap}
\end{figure}

\begin{table*}
\begin{center}
\begin{tabular}{||l||c|c|c|c|c||} \hline
system & central mass (in $M_\odot$)   & $\alpha$-parameter    & radii                 & accretion rate (in $M_\odot$/yr) & number of runs\\ \hline
AGN     & $10^8$               & 0.01 and 0.1          & $20-10^4 R_{\rm S}$ &  $10^{-4} - 0.4$ &  $9372$\\
DN      & 0.61                 & 0.01 and 0.1          & $10^9-10^{11}$ cm     & $10^{-12} - 10^{-8}$ & $7742$ \\
YSO     & 1                    & 0.001 and 0.01        & $0.05-50$ UA   & $10^{-9} - 10^{-5}$ & $11534$\\ \hline
\end{tabular}
\end{center}
\caption{Domain of calculation, depending on the system ($R_{\rm S}$ is the Schwarzschild radius of the black hole).  In all cases, the $(\log \dot{M}, \log R)$-plane is explored with constant logarithmic steps corresponding to $\frac{\Delta \dot{M}}{\dot{M}}=\frac{1}{8}$ and $\frac{\Delta R}{R}=\frac{1}{10}$ for the accretion rate and radius respectively. The last column gives the number of runs performed for the comparisons.}
\label{tab:doc}
\end{table*}

Figure \ref{courbe} displays $h(R)$ and $T_{\rm mid}(R)$ obtained from the two
 models for parameter values $(M,\dot{M},\alpha)$ typical of a dwarf nova (DN) disc, a young stellar object (YSO) disc
 and a disc in an active galactic nucleus (AGN). We see that the VE-model always gives a slightly thicker and hotter disc than the VA-model, but the agreement is globally rather good. The other quantities, not shown here, compare very well too.

In order to test the
  sensitivity of these results to the values of the input parameter values, we have performed a
 series of computations by varying the accretion rate and the
 value of $\alpha$, for any of the above three systems. This amounts to about $29 000$ runs in total. Since the one
 layer model yields a multivalued solution (MVS), not
 present in the VE-model (Cannizzo, 1992), the comparison
 has been restricted to the hot, inner and cold,
 outer branches located on both sides (see Fig. \ref{courbe}). The approximate location of the MVS region will be specified later.

\begin{table}
\begin{center}
\begin{tabular}{||l||*{2}{c|}|}
  \hline
   \multicolumn{3}{||c||}{$\alpha = 0.1$} \\
  \hline
   & \bfseries hot, inner branch & \bfseries cold, outer branch \\
  \hline
   $\epsilon (T_{\rm mid})$                & $1.0 - 1.2$ 
                                           & $1.1 - 1.2$ \\
  \hline
   $\epsilon (h)$                          & $1.2 - 2.8$  
                                           & $2.2 - 3.9$ \\ 
  \hline
   $\epsilon (\rho_{\rm mid})$             & $0.8 - 3$   
                                           & $1.0 - 2.8$ \\
  \hline
   $\epsilon (\Sigma_{\rm t})$             & $1.1 - 2.2$ 
                                           & $1.3 - 1.6$ \\
  \hline
   $\epsilon (\tau_{\rm t})$               & $1.0 - 2.3$   
                                           & $0.6 - 2.4$ \\
  \hline
   \hline 
   \multicolumn{3}{||c||}{$\alpha = 0.01$} \\
  \hline
   & \bfseries hot, inner branch & \bfseries cold, outer branch \\
  \hline
   $\epsilon (T_{\rm mid})$                & $1.0 - 1.2$ 
                                           & $1.1 - 1.2$ \\
  \hline
   $\epsilon (h)$                          & $1.8 - 3.8$  
                                           & $2.5 - 4.0$ \\ 
  \hline
   $\epsilon (\rho_{\rm mid})$             & $0.8 - 1.5$   
                                           & $1.0 - 1.3$ \\
  \hline
   $\epsilon (\Sigma_{\rm t})$             & $0.9 - 1.7$ 
                                           & $1.0 - 1.3$ \\
  \hline
   $\epsilon (\tau_{\rm t})$               & $0.9 - 1.6$   
                                           & $0.6 - 1.6$ \\
  \hline 
\end{tabular}
\end{center}
\caption{Extreme values of $\epsilon$ for
         an AGN disc, and for two different values of the $\alpha$-parameter, outside the MVS region (see Fig. \ref{courbe} and Table \ref{tab:doc} for the domain of computation).}
\label{tab:AGN1}
\end{table}

Figure \ref{fig:htrmap} displays iso-values of the quantity $\epsilon(T_{\rm mid})$ in the
$(\log \dot{M}, \log R)-$plane in the AGN case, for the canonical values $M=10^8$ M$_\odot$ and
$\alpha=0.1$. From this data grid, the maximum and minimum deviations on the temperature ratio can easily be extracted. So, we find $1 \lesssim \epsilon(T_{\rm mid}) \lesssim 1.2$ over the radial domain and for accretion rates indicated in Table {\ref{tab:doc}}. We have repeated this procedure for the other four key quantities $h$, $\rho_{\rm mid}$, $\Sigma_{\rm t}$ and $\tau_{\rm t}$, and for $\alpha=0.01$ as well. Table \ref{tab:AGN1} summarizes the results. We notice that the agreement between the two approaches is excellent, indeed. The midplane temperature is slightly underestimated with the VA-model, by a factor never exceeding $20\%$ (in the cold outer disc) for any $\alpha$, which is remarkable given the
 crudeness of this model. The other quantities are a little bit more
 affected by the vertical averaging: we have a mean factor $\approx 2$ on the central density
 and $\approx 1.5$ on the optical thickness and surface density. The
 largest departures are observed on the disc thickness, with a factor reaching
  $\approx 4$ in the cold, outer disc. The sensitivity to the value of the
 $\alpha$-viscosity parameter is weak, even null for the temperature. However, the trend is that the deviations seem smaller with a smaller value of the $\alpha$-parameter, except for $h$.

\begin{table}
\begin{center}
\begin{tabular}{||l||*{2}{c|}|}
  \hline
   \multicolumn{3}{||c||}{$\alpha = 0.01$} \\
  \hline
    & \bfseries hot, inner branch & \bfseries cold, outer branch \\
  \hline
   $\epsilon (T_{\rm mid})$                & $0.9 - 1.1$ 
                                           & $1.0 - 1.3$ \\
  \hline
   $\epsilon (h)$                          & $1.2 - 1.9$  
                                           & $1.0 - 3.3$ \\ 
  \hline
   $\epsilon (\rho_{\rm mid})$             & $1.2 - 1.6$   
                                           & $0.4 - 1.4$ \\
  \hline
   $\epsilon (\Sigma_{\rm t})$             & $1.7 - 3.4$ 
                                           & $0.9 - 1.8$ \\
  \hline
   $\epsilon (\tau_{\rm t})$               & $1.4 - 2.1$   
                                           & $0.3 - 1.9$ \\
  \hline
  \hline
   \multicolumn{3}{||c||}{$\alpha = 0.001$} \\
  \hline
    & \bfseries hot, inner branch & \bfseries cold, outer branch \\
  \hline
   $\epsilon (T_{\rm mid})$                & $0.9 - 1.1$ 
                                           & $1.0 - 1.2$ \\
  \hline
   $\epsilon (h)$                          & $1.1 - 2.6$  
                                           & $1.5 - 4.0$ \\ 
  \hline
   $\epsilon (\rho_{\rm mid})$             & $1.1 - 2.1$   
                                           & $1.0 - 1.4$ \\
  \hline
   $\epsilon (\Sigma_{\rm t})$             & $1.5 - 2.9$ 
                                           & $1.3 - 1.8$ \\
  \hline
   $\epsilon (\tau_{\rm t})$               & $1.3 - 2.4$   
                                           & $0.7 - 2.0$ \\
\hline
\end{tabular}
\end{center}
\caption{Same as for Table \ref{tab:AGN1} but in the YSO case.}
\label{tab:YSO}
\end{table}

\begin{table}
\begin{center}
\begin{tabular}{||l||*{2}{c|}|}
  \hline
   \multicolumn{3}{||c||}{$\alpha = 0.1$} \\
  \hline
    & \bfseries hot, inner branch & \bfseries cold, outer branch \\
  \hline
   $\epsilon (T_{\rm mid})$                & $1.1$ 
                                           & $1.1-1.7$ \\
  \hline
   $\epsilon (h)$                          & $1.0-2.8$  
                                           & $1.8-3.8$ \\ 
  \hline
   $\epsilon (\rho_{\rm mid})$             & $1.1-2.6$   
                                           & $0.3-1.3$ \\
  \hline
   $\epsilon (\Sigma_{\rm t})$             & $1.5-3.2$ 
                                           & $0.6-1.7$ \\
  \hline
   $\epsilon (\tau_{\rm t})$               & $1.3-2.9$   
                                           & $0.8-6.0$ \\
  \hline
  \hline
   \multicolumn{3}{||c||}{$\alpha = 0.01$} \\
  \hline
    & \bfseries hot, inner branch & \bfseries cold, outer branch \\
  \hline
  \hline
   $\epsilon (T_{\rm mid})$                & $1.1$ 
                                           & $1.1$ \\
  \hline
   $\epsilon (h)$                          & $1.2 - 2.9$  
                                           & $2.6 - 2.9$ \\ 
  \hline
   $\epsilon (\rho_{\rm mid})$             & $1.2 - 1.4$   
                                           & $0.9 - 1.6$ \\
  \hline
   $\epsilon (\Sigma_{\rm t})$             & $1.5 - 1.8$ 
                                           & $1.4 - 1.5$ \\
  \hline
   $\epsilon (\tau_{\rm t})$               & $1.3 - 1.8$  
                                           & $1.2 - 1.8$ \\
  \hline
\end{tabular}
\end{center}
\caption{Same as for Table \ref{tab:AGN1} but in DN case.}
\label{tab:CV}
\end{table}

We have carried out similar computations in the YSO and DN cases. The domains of calculation, specific to each system (Gullbring et al., 1998; Horne, 1998), are listed in Table \ref{tab:doc} and the results are
summarized in Tables \ref{tab:YSO} and \ref{tab:CV} respectively. Globally, the conclusions drawn in the AGN case
still hold. The best agreement between the two models is observed in the DN case. In particular, the temperature ratio is constant ($\approx 1.1$) throughout, whatever the value of $\alpha$. However, for $\alpha=0.1$ and for the lowest accretion rates examined, the optical thickness becomes close to unity in the outer disc and so the optically thick approximation underlying the two models becomes doubtful. This produces a specially large ratio $\epsilon(T_{\rm mid})  \sim 1.7$.

We point out that the deviations between the two models are largest in the vicinity of the MVS region, on the hot branch side (see Fig. \ref{courbe}). Since our radial resolution (set to $\frac{\Delta R}{R}=0.1$ in the present calculations) is finite, it is expected that the ``real'', maximum deviations (i.e. the deviations resulting from an infinite resolution) are probably slightly different to those reported in the tables.

Gathering the results obtained for the three systems considered here, we find that the MVS region which more or less encompasses the thermally unstable regime where hydrogen recombines (Meyer \& Meyer-Hofmeister, 1982) is located at
\begin{equation}
 R_{\rm MVS} \simeq  ( 1 \pm 0.6) \times 10^{15}  \alpha^{-0.2} M_0^{0.34} \dot{M}_0^{0.44} \qquad {\rm cm}
\label{eq:mvsregion}
\end{equation}

where the mass $M_0$ and accretion rate $\dot{M}_0$ are expressed in M$_\odot$ and M$_\odot$/yr respectively. The universal character of this expression can easily be understood from power law solutions of the VA-model (Hur\'e, 1998), given that the MVS region (see Fig. \ref{fig:htrmap}) is roughly bounded by isotherms.

\section{Concluding remarks}

Given the ranges of parameter values (masses, accretion rates and viscosity parameter values) considered in this study, we can conclude that the vertically averaged model of steady state Keplerian $\alpha$-discs is remarkably reliable, since it provides solutions very close to those obtained from the vertically explicit version. The midplane temperature appears as the quantity least sensitive to the vertical averaging (with a $30\%$ deviation only) whereas the disc thickness seems the most affected (a factor $\sim 4$ in some cases). Problems dealing with the disc irradiation (even with a global energy budget) should therefore account for vertical stratification since the disc response is sensitive to the local flaring angle and radial gradient (Kenyon \& Hartmann, 1987).

This investigation demonstrates that the vertically averaged approach should be sufficient for many applications (e.g. Pringle, 1997; Burderi, King \& Szuszkiewicz, 1998; Collin \& Hur\'e, 1999), in particular if the level of accuracy required on the disc structure is low or moderate. Note however that the tables given here can be used to ``correct'' the vertically averaged solution if a better precision is necessary. More basically, these tables show the typical error on the knowledge of the disc structure, that is, some methodology effect. It is interesting, but comforting, to see that the model is much more sensitive to the physics it contains (like turbulent viscosity, convection, opacities, etc.) than to the method used to derive the disc structure.

\begin{acknowledgements}
I thank M. Mouchet for reading the manuscript, D. Richard, and the referee, Dr John K. Cannizzo, for interesting remarks, specially regarding convection.

\end{acknowledgements}

\end{document}